\documentclass[letterpaper,twocolumn,10pt]{article}
\usepackage{usenix2019_v3}

\usepackage{tikz}
\usepackage{amsmath}
\usepackage{xcolor}

\usepackage[english]{babel}
\usepackage{blindtext}

\usepackage{caption}
\usepackage{subcaption}
\usepackage{booktabs}
\usepackage{tabularx}
\usepackage{makecell}
\usepackage{enumitem}
\usepackage{graphics}
\usepackage{multirow}
\usepackage{array}
\usepackage[
  separate-uncertainty = true,
  multi-part-units = repeat
]{siunitx}

\newcommand{\systemName}[0]{EnCoR\xspace}

\makeatletter
\def\@makefnmark{%
  \leavevmode
  \raise.9ex\hbox{\fontsize\sf@size\z@\normalfont\tiny\@thefnmark}}
\makeatother

\begin{document}

%don't want date printed
\date{}

% make title bold and 14 pt font (Latex default is non-bold, 16 pt)
\title{\Large \bf \systemName: An end-to-end architecture for simplifying cellular networks}

%for single author (just remove % characters)
\author{
{\rm Wesley Woo}\\
Virginia Tech
\and
{\rm Zhuowei Wen}\\
Virginia Tech
\and
{\rm Monniiesh Velmurugan}\\
Virginia Tech
\and
{\rm Richard Raad}\\
Virginia Tech
\and
{\rm Sylvia Ratnasamy}\\
UC Berkeley
\and
{\rm Scott Shenker}\\
UC Berkeley/ICSI
\and
{\rm Shaddi Hasan}\\
Virginia Tech
} % end author

\maketitle

\begin{abstract}
Since their creation, cellular networks have made in-network mobility support a key feature of their service model.
While this approach provides seamless connectivity for legacy traffic, it has the side effects of inflating end-user latency and increasing complexity and operational overhead for operators.
Yet modern applications and transport protocols are increasingly mobility tolerant, prompting us to revisit the assumption that mobility must be provided as an in-network service.

In this paper, we propose \systemName (End-to-End Core and RAN), a deployable cellular network architecture that removes mobility from the core entirely.
Leveraging end-to-end mobility, \systemName eliminates tunnel-based IP anchoring while preserving compatibility with existing authentication, charging, and QoS techniques.
We demonstrate that \systemName works with unmodified phones while providing equivalent performance as traditional LTE networks for real applications including video and voice calling and video streaming.
We show that \systemName not only allows network operators to reduce end to end latency, but can also reduce the capital cost of providing low latency service to users by more than 90\% compared to 3GPP networks, based on cost estimates for cellular network core and border router infrastructure provided by the FCC.
Finally, we demonstrate that these gains are achieved while reducing the amount of overall handover control messaging, allowing the \systemName core network to handle a greater number of mobility handover events than an LTE core under identical hardware constraints, achieving a 2.6x lower handover latency under load.

\end{abstract}
\setcounter{page}{1}

\section{Introduction}
Today's Internet ecosystem is defined by endpoints in motion.
Indeed, mobile phones are the dominant means by which humans connect to the Internet~\cite{ITU2023}, with more than half of the world's population accessing the Internet via mobile networks~\cite{GSMA2023}.
Emerging use cases, such as XR and autonomous vehicles, assume mobile end devices and place stringent performance demands on networks under mobility~\cite{schafhalter2025bandwidthallocationcloudaugmentedautonomous}.
This has made \emph{mobility} a critical network service despite not being part of the Internet's original service model~\cite{clark1988design}.

Today’s cellular networks represent the state of the art for mobility support within a single network, with mobility being core to their service model.
For example, modern 5G networks aim to support scenarios such as devices moving on high speed rail at over 500 km/h and sub-millisecond switching between radios to provide so-called "ultra-reliable low latency communication" (URLLC)~\cite{ericsson2020reducing}.

However, even with these advances, the overall \textit{architectural} approach cellular networks take to support mobility remains largely unchanged from the earliest mobile Internet service provided by 2G networks developed in the 1990s.
These networks provided access to a very different Internet ecosystem compared to today's.
TCP was the dominant transport layer protocol, mobile Internet speeds (for the few who had access to those services) were measured in kbps, smartphones did not exist, and CDNs were in their infancy.
Circuit-switched voice calls were the most important application on these early mobile networks, which rely on centralized core networks for call switching.
As such, the first support for mobile Internet traffic took a similar approach, tunneling traffic to a central aggregation point in the network core that serves as an anchor for users' IP addresses.
Today's cellular networks continue to provide mobility as a fundamental, critical part of their service model by tunneling user traffic from radio access network to an aggregation point (the P-GW in LTE networks or UPF in 5G networks) in the network core.

This approach has drawbacks in today's Internet.
For users, it inflates end-to-end latency by detouring traffic through a small number of mobility anchors in operators' core networks.
Novel 5G architectures such as Multi-access Edge Computing (MEC) promise to address this by moving traffic anchoring closer to the edge — but because these anchors remain stateful middleboxes on the datapath, deploying more of them closer to users means proportionally more state to manage, more control plane traffic, and higher deployment costs.
In other words, the latency benefits of MEC are fundamentally at odds with the scalability challenges the architecture imposes on operators.
Beyond performance and cost, mobility as an in-network service also complicates operations in the increasingly heterogeneous networks operators deploy: a network operator seeking to deploy 5G alongside an existing LTE network must ensure that their combined network can handle mobility scenarios across disparate network implementations.
Similar complications arise when roaming between operators. 

In this paper, we ask how a cellular network might be designed if we could instead rely upon \emph{end-to-end} mobility.
The state of the modern Internet encourages us to take a broad view of what end-to-end mobility can look like.
The \emph{transport} layer has become increasingly mobility-tolerant: there have been many proposals for transport layer support for host mobility~\cite{snoeren2000end,raiciu2011opportunistic,aydin2003cellular,honda2007smsctp,koh2004msctp}, 
and protocols such as QUIC and HTTP/3 that provide explicit support for connection migration are widely adopted~\cite{facebook_quic_adoption_2020,langley2017quic,madariaga2020analyzing}.
At the same time, today's \emph{applications} are themselves mobility tolerant: users regularly move between WiFi and cellular networks, requiring application developers to handle connection interruptions as a common case.
For example, video applications maintain buffers to handle short periods of disconnection, and social media applications perform story ranking on mobile devices to improve performance on poor networks~\cite{petrescu2016client}.

Projecting these application and transport layer trends forward, and in light of increasing demands to reduce cellular latency~\cite{3dincities_low_latency_driving, ericsson_low_latency_telesurgery}, we argue that the time is right to revisit the assumption that mobility must be an in-network service.
Our argument follows directly from the end-to-end principle~\cite{saltzer1984end}, as an increasing range of modern applications do not require support from the cellular core to handle mobility, and since providing this service incurs costs for both applications (in the form of inflated latency and tunneling overhead) and operators (in the form of operational complexity).
As we will show, not only could such an approach address challenges faced by today’s cellular networks, such a design is actually achievable given the realities of the modern internet.

To do this, we propose \systemName (\emph{End-to-End Core and RAN}), a cellular network architecture that relies upon end-to-end mobility support.
\systemName takes advantage of end-to-end mobility by decomposing the existing core network into stateful and stateless components.
Stateful functions that allow a cellular operator to maintain control of their network (such as authentication and charging) remain centralized, whereas stateless functions involved in the critical path, such as mobility, can be distributed via lightweight services around the edge of the network.
In doing so, \systemName reduces core network complexity, improves network scalability, and reduces end user latency.

We implement \systemName and show that removing in-network mobility enables large cost and latency improvements, does not break modern applications, scales better under mobility load, and can be realized today without changes to existing phones.
Specifically, by routing traffic directly at the network edge, \systemName achieves the lowest end-to-end latency achievable under existing cellular architectures at over 90\% lower cost.
Using both custom QUIC-based applications and unmodified commercial applications on real devices, we show that existing applications continue to operate correctly and without measurable performance degradation in the absence of in-network mobility support. 
By simplifying the control plane, \systemName reduces mobility-related control overhead and achieves up to 2.6x faster handover completion under load compared to LTE on identical hardware.
Finally, because \systemName requires no change to existing phones or radio interfaces, it is deployable alongside existing cellular infrastructure.

\textbf{Evaluation approach.}
This work contends that adopting an end-to-end mobility paradigm for cellular networks both reduces operational complexity and improves end-to-end latency and control plane load through the core during handover while matching the performance of traditional mobile networks.
Evaluating these claims is challenging due to the scale and complexity of modern cellular networks.
Operational complexity depends on an operator's organizational practices and cost structure, which can vary widely and are difficult to measure outside an operational context.
Similarly, realistic evaluation of user and control plane traffic patterns is hindered by the sheer scale of these networks, and even appropriately granular traffic traces are hard to obtain outside of a mobile operator due to privacy, security, and scale constraints.
Given these constraints, this paper approaches evaluation of \systemName by (1) benchmarking mobility impact on individual applications using an open source hardware testbed, (2) estimating end-to-end latency gains through simulation of a nation-scale cellular network built from real-world deployment data, and (3) reporting control plane load handling under scale by approximating production-scale resource contention using
throttled hardware.

\textbf{Roadmap.} We begin with a review of related work (\S\ref{s:related}) and background on the state of mobility in today's cellular networks (\S\ref{s:background}). We then outline the design and prototype implementation of \systemName (\S\ref{s:design} and \S\ref{s:implementation}) before evaluating our system (\S\ref{s:evaluation}). We end with discussion and conclusion (\S\ref{s:discussion} and \S\ref{s:conclusion}).
This work does not raise any ethical issues.

\section{Related Work}
\label{s:related}

\textbf{Mobility in NextG networks.}
Mobility management remains both a core function and a significant source of complexity in today's 5G cellular networks \cite{hassan2022sigcomm, noh2020ieee_wc}.
Recent work explores how current mobility handling procedures can be improved to support seamless mobility, such as through the use of machine learning algorithms for triggering handover decisions~\cite{5gmob_mollel2021survey, 5gmob_rl_yajnanarayana2020} or novel network architectures to support high-speed mobility scenarios~\cite{high5gmob_fan2016}.
These approaches improve mobility performance, but maintain the architectural approach long used in cellular networks where 
mobility decisions are made in radio access networks and a centralized data plane anchors IP addresses for mobile devices.
Similar work in distributed cellular network architectures have taken similar approaches to re-imagining network architecture~\cite{dauth, dlte,hasan2023building}, in some cases even exploring end host-driven mobility to achieve their designs~\cite{cellbricks}.
Our work is distinct from this prior work in that it positions end host mobility as a central design feature, rather than as a means to achieving some other design goal (e.g. democratizing cellular access~\cite{cellbricks}).

\textbf{Low-latency in 5G.}
In 5G, Multi-access Edge Computing (MEC) seeks to address the latency performance bottleneck of a centralized user plane both by terminating the in-network data path at network edge with dense User Plane Function (UPF) deployments and by supporting intelligent routing to application server deployments with 5G Uplink Classifiers~\cite{5gmec_survey, 5g_ulcl}.  
While these approaches in theory support the complete elimination of user plane tromboning through the core network, realizing this performance characteristic requires costly deployments (in the worst case, one UPF for every base station) and introduces significant additional control messaging to support changing serving UPFs during handover in both make-before-break or break-before-make Service and Session Continuity modes (SSC Modes 2 and 3)~\cite{3gpp23502ssc_handover, 3gpp23502session_management}.
Our work explores the complete removal of mobility anchors in 3GPP cellular network architectures, realizing low-latency performance without drastic increases in control plane traffic or deployment cost.
We specifically demonstrate this advantage over MEC through simulation in \ref{s:eval:latency}. 

\textbf{End-to-end mobility at the transport layer.}
The idea of mobility support as an end-to-end service is not new.
Previous literature has explored providing mobility support at the application or session layer~\cite{sessionmobility, perkins1997mobile} and at the transport layer~\cite{mobilityfirst, mptcpmobility, freeze_tcp}.
Indeed, in some cases end-to-end mobility performance can slightly exceed the performance of in-network mobility \cite{cellbricks}.
This work identifies QUIC as a \emph{widely deployable} end-to-end mobility architecture and outlines how QUIC's end-to-end mobility capabilities support an improved cellular network design.
While others have considered QUIC in mobility scenarios, including proactive connection migration~\cite{tan2020proactive} to pick the optimal network path in a multi-network scenario and QUIC-centric clean-slate Internet architectures~\cite{tauqeer2024use}, our work explores how QUIC can be used on devices without multi-network access in \textit{today's Internet}.

\section{Background}
\label{s:background}
In this section, we provide a background on modern cellular network architecture, beginning with an overview of cellular network design before shifting focus to elements and procedures that support mobility.
While the specific decomposition of functionality within cellular networks varies across generations, the architectural approach to mobility support is largely consistent across all generations of 3GPP\footnote{3GPP is the standards body for all widely-deployed modern cellular networks, including 2G GSM, 3G UMTS, 4G LTE and 5G NR networks.} cellular networks.
Where we refer to specific generation details we refer to LTE, as this remains the most common cellular \emph{core} network architecture in deployment today, even for networks that use 5G radio infrastructure.

\textbf{Overview of 3GPP Mobile Networks.}
3GPP networks divide functionality between a distributed \emph{radio access network (RAN)} and a logically centralized \emph{core network}.
The RAN consists of base stations and radio equipment on towers, while the core implements network functions that support the entire network.
These include \emph{control plane} functions that manage subscriber state, authentication, and mobility, as well as \emph{user plane} functions that carry user traffic between mobile devices and the Internet.

In modern networks, the RAN and the core connect via an IP fabric.
User traffic is tunneled from the user's device (the "UE") via the RAN to the core network as an overlay on top of this IP fabric, with tunnel configuration managed by the core network's control plane.
The core network's user plane anchors the UE's IP address and updates tunnel routing state as the user moves; from the user's perspective, her IP address remains fixed, even as she moves between base stations.

To illustrate how these components interact, consider the procedure a UE performs to connect to an LTE network.
The UE establishes a radio link to a base station (the eNodeB), which forwards the connection request to the core network.
The UE and the core perform mutual authentication using shared keys stored in the UE's SIM and the core network's subscriber database function (the HSS).
The core network's mobility function (the MME) derives session keys to encrypt the radio link, distributes these to the base station and the UE, and then triggers the core network's user plane functions (the S-GW and P-GW) to create an IP session for the UE.
The P-GW allocates an IP address and, with the help of other core network components, sets up tunnels to route user traffic between the UE and the Internet.

The key tasks during attach are mutual authentication, distribution of session keys, and configuration of the user plane's tunnel overlay.
Across all 3GPP generations, \textit{this user plane anchors the UE's IP address in one central physical location for the duration of the UE's connectivity session, even as the UE physically moves around the network}.
This detail is key to understanding the design and limitations of mobility support in 3GPP networks.

\textbf{Mobility in 3GPP Networks.}
Mobility in 3GPP networks is realized through \textbf{handover}, which moves a UE from one base station to another.
Handover is triggered by the radio access network when it detects a UE would receive better service from a different base station than the one it is currently attached to.
The two base stations involved first prepare the handover by exchanging physical layer parameters with each other that are forwarded to the UE to facilitate a fast physical layer reconnection.
These include the target base station's identifiers, radio frequency, channel configuration, and timing information.
In parallel, the RAN and the core coordinate to construct a new set of overlay tunnels for the UE, derive new session keys for the new base station, and buffer incoming traffic while the UE is changing base stations.
Once this preparation is complete, the RAN directs the UE to handover to the new base station.

This three step process -- initiation, preparation, and execution -- can occur billions of times per day in a mobile network~\cite{through_telco_lens}.
Because each handover involves multiple elements of the RAN and the core network, control plane signaling associated with handover comprises a significant portion of total control plane load on the cellular network.

\textbf{Drawbacks to 3GPP Mobility.}
3GPP's approach to mobility assumes that IP address changes need to be hidden from higher layers.
In other words, it assumes that the transport layer and application layers are not mobility tolerant, which makes in-network mobility a \textbf{correctness requirement}.

This approach presents several drawbacks.
First, these stateful functions must handle mobility events and associated control plane signaling at a rate proportional to users' movement in the network.
Second, routing traffic via tunnels incurs tunnel overheads and requires detouring through mobility anchors, inflating latency; this is particularly unfortunate if the user is otherwise close to a edge server that could serve the request as traffic must still hairpin through the anchor in the cellular core.
Finally, and more subtly, handover between heterogeneous systems requires all  network elements to work together, making testing and deployment fragile.
This is precisely the challenge faced by operators upgrading their network from one generation to the next: during the transition, handover between generations is common~\cite{hassan2022sigcomm}, and a single LTE to 5G handover would involve coordination between the MME, P-GW, AMF, SMF, and UPF, in addition to each radio and other supporting elements.
These costs are incurred by all users and applications, even those that do not require this level of mobility support.
\section{Design}
\label{s:design}
We propose \systemName, a deployable cellular network architecture that benefits from end-to-end mobility while achieving:

\begin{itemize}[noitemsep,topsep=0pt]
    \item \textbf{Performance:} Provides end-user performance and latency equivalent to or exceeding today's networks, including in mobility scenarios.
    \item \textbf{Scalability:} Supports national-scale networks, equivalent to today's LTE and 5G networks.
    \item \textbf{Deployability:} Re-uses as much of the existing cellular ecosystem as possible. 
    \item \textbf{Compatibility with MNO business needs:} Supports the same authentication, charging, per-subscriber policy, and security model MNOs require today.
\end{itemize}

\begin{figure}
  \includegraphics[width=0.9\columnwidth]{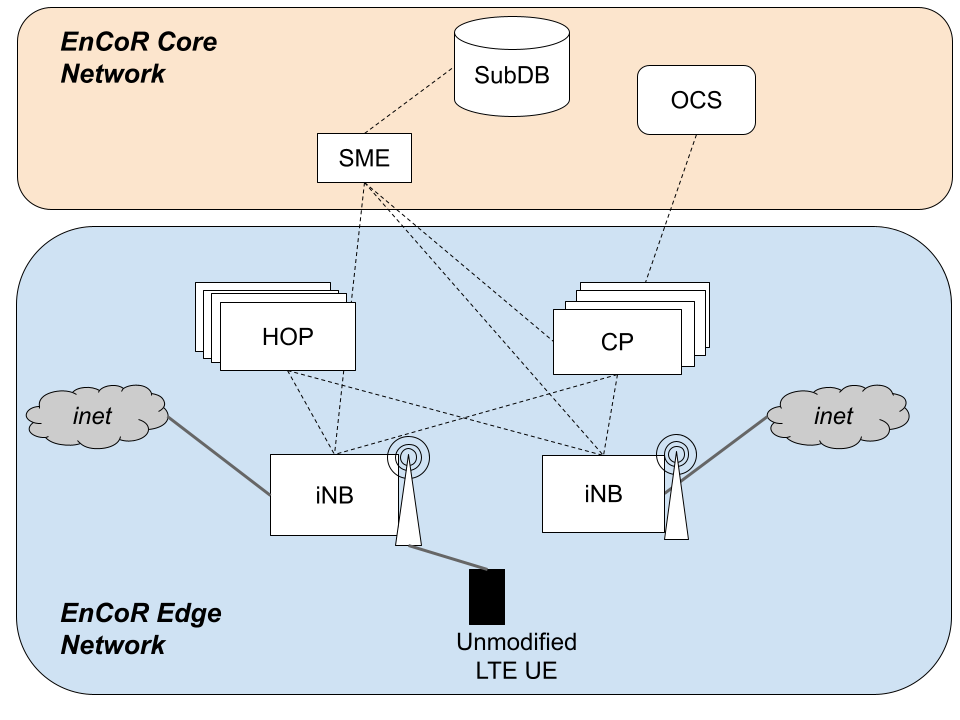}
  \caption{\systemName architecture}
  \label{fig:system_arch}
\end{figure}

Fig.~\ref{fig:system_arch} depicts the \systemName architecture.
Like existing 3GGP mobile networks, \systemName is composed of a core network and and edge network, both overlayed on top of an operator's existing IP network.
Further, \systemName preserves the 3GPP radio interface used between existing phones and base stations, allowing these to be used unchanged, as these represent significant investments by network operators and users.

From here, \systemName makes a key point of departure from traditional 3GPP designs: we remove mobility anchoring of UE IP addresses from the core.
Examining the state that remains in the core, we observe that only a part of it is truly stateful and the rest is ephemeral state only relevant to a particular session.
In addition, only the ephemeral state is on the critical path for user traffic.
Hence, we choose to terminate the user plane at the network edge, effectively eliminating the need for centralized elements of the current cellular user plane and the control plane overhead associated with coordinating them.
This removes the latency penalty incurred by today's 3GPP networks due to central detouring and removes all centralized stateful middleboxes from the critical path and allows us to decompose \systemName's control plane into a compact, stateful core and a lightweight, stateless edge that collectively support a largely unmodified RAN.

\subsection{Overview}

\systemName divides functionality between the edge user plane, a lightweight stateless edge control plane, and a stateful centralized core control plane.

\paragraph{Edge user plane.} User data in \systemName flows from UE to base station over the same radio interface as LTE and 5G networks, allowing existing 3GPP radio infrastructure and handsets to continue working without modifications.
However, \systemName terminates the user plane at the network edge by routing traffic to external IP networks from base stations.
To achieve this, \systemName extends the traditional eNB/gNB base station to take on the role of connecting the network to external networks (the role of the P-GW in LTE); we term this "extended base station" network element the \emph{iNodeB (iNB)}.

\paragraph{Stateful central control plane.} As with 3GPP mobile networks, \systemName control plane functions requiring long-lived persistent user state are logically centralized in a compact core control plane.
This core handles critical functions such as user authentication, provisioning, and charging.
To support authentication and provisioning, \systemName relies on a Security Management Entity (SME) and a Subscriber Database (SubDB).
To support user charging, \systemName adopts standard quota-based online charging to allow operators to perform usage-based billing via an Online Charging System (OCS). 

\paragraph{Stateless edge control plane.} 
Finally, \systemName adds a lightweight edge control plane layer alongside the RAN in the edge sub-network to handle latency sensitive control plane operations that require only ephemeral session state.
This edge control plane consists of two types of network element: Handover Proxies (HOPs) and Charging Proxies (CPs).
HOPs facilitate efficient mobility by allowing iNBs to exchange control plane messages related to handover that bypass the central control plane (Section~\ref{s:ho_procedure}).
Because HOPs are stateless, they can be replicated widely throughout the network to handle signaling load and reduce signaling latency between iNBs.
CPs play a similar role for charging state, caching subscriber session identifiers, usage counters, and quotas, and batching updates to the centralized OCS.

Together, this decoupling provides the foundation for a scalable, deployable cellular network that targets the same use cases as today's 3GPP networks. 
We now turn to the detailed design of \systemName.

\subsection{Edge user plane}
\label{s:user_plane}
In \systemName, the iNodeB comprises the entirety of the cellular user plane and passes user traffic directly to the mobile network operator's underlying IP network.
Specifically, the iNB bridges user traffic between the cellular network and the broader Internet, applies Quality of Service (QoS) markings where required, and enforces usage quotas.
Like other RAN elements in today's 3GPP networks, the iNB also retains responsibility for deciding when to move a user's device from one RAN element to another, preserving operator control over load balancing across cells and user quality of experience.

\textbf{IP address assignment and stateless NAT.} A key function of the user plane is to assign an IP address to connected UEs.
In 3GPP networks, this occurs when the UE initially connects to the network, with a global IP address assigned to each UE.
A naive approach for address assignment in \systemName would be for each iNB to assign a new IP address to each connected UE.
This would make mobility fully transparent to end-host applications, but it would require each UE to learn its new IP address from every iNB it connects to before it could continue sending data.
This would require either an additional round-trip between UE and iNB on each mobility event, or sending additional IP address state as part of the handover procedure, which would incur additional signaling overhead and break compatibility with existing handsets.

Instead, we assign a private IP address to every UE upon initial connection, and then have each iNB perform network address translation for all traffic.
Although traditional stateful NAT is sufficient, to avoid overhead of large NAT tables on the iNB, we propose a \emph{stateless NAT} approach, inspired by identifier-locator address schemes like LISP~\cite{rfc6830}, ILNP~\cite{atkinson2009ilnp}, and ILA~\cite{herbert-intarea-ila-01}.
First, each UE is assigned a unique private IPv6~\cite{rfc4193} address upon initial connection (e.g., in the reserved \texttt{fc00::/7} subnet), where the lower 64 bits are used to uniquely identify the device within the operator's network.
The precise mechanism for assigning these unique addresses can vary: these could be assigned by the central control plane, be pre-assigned in each device's SIM card, or even be derived deterministically based on the SIM card's unique ID.
The key is to avoid identity collisions as UEs move across the network.
Then, each iNB in the operator's network is assigned a publicly routeable IPv6 subnet in the upper 64 bits.
To perform address translation, the iNB need only replace the upper half of the source address of packets from each UE with its own subnet, and perform the reverse for downlink traffic.

From the UE's perspective, its IP address never changes, but hosts that it is communicating with on the Internet will see the updated public IP address from the iNB.
Put another way, this allows the iNB to assign IP addresses \emph{on demand and transparently} to the UE.
This approach allows us to re-use existing handover signaling procedures while avoiding additional control signaling or complex NAT tables. %\sr{neat!}

\textbf{QoS and Quota Enforcement.}
Quality of Service enforcement is a key use case in 3GPP mobile networks that enables network operators to offer differentiated service to users on a per application basis.
Today, cellular networks enforce quality of service using standardized 3GPP QoS Class Identifiers (QCIs) to differentiate traffic.
QCI configurations are applied on a per-user basis and are captured in the network's central configuration (our SubDB).
For downlink traffic (towards the UE), iNBs enforce QCI scheduling and prioritization using functionality already implemented by RAN equipment.
iNBs apply DiffServ markings and prioritize traffic on uplink traffic (from the UE towards the Internet) to ensure appropriate handling on the operator's IP network.
Because QCIs are communicated between RAN elements during handover, we do not require coordination with the core to apply QoS rules.

Service quota enforcement is handled similarly.
In general, mobile operators apply service quotas on the basis of volume (e.g., bytes transferred), time duration, of throughput.
As with QoS configuration, time and throughput based usage policies are typically stored in the SubDB and applied during initial connection.
Throughput-based policies are typically enforced by radio scheduling.
Volume based charging typically involves the use of standardized online charging systems that allocates a quota of traffic volume to a user plane element which in turn enforces limits.
These are all simple policies to support in the data path, as they require only per-user counters and the ability to cut off service to a user; we will discuss how \systemName handles the control plane associated with this in Section~\ref{s:charging_proxies}.
Prior work has shown such policies and even more sophisticated ones involving content classification can be easily supported on a distributed user plane~\cite{hasan2023building}; a production implementation of \systemName could adopt similar techniques.

\textbf{Overhead.}
Our proposed extensions to iNBs are modest compared to the workloads typical RAN elements already handle.
Most production eNodeBs are split between one or more radio units (RUs) connected to a single baseband unit (BBU).
OpenRAN~\cite{openran}, for example, defines a number of functional splits that isolate specialized and frequency-specific hardware (RU) from general purpose hardware running various layers of the BBU.
As the BBU is already provisioned to handle encoding and decoding of the physical layer of the RU, adding stateless NAT and per-user counters for managing charging quotas is not a significant increase in workload.

\subsection{Minimal Central Control Plane}
\label{s:central_core}
By terminating the user plane at the edge, the responsibilities of the control plane become restricted to authentication, security, and charging -- all services that depend on a notion of global, persistent state.

\textbf{Components.}
In \systemName, the key central control plane functions are the Subscriber Database (SubDB), which is responsible for storing subscriber identity, cryptographic keys, and subscription information (such as QoS policy); and the Security Management Entity (SME), which handles the basic lifecycle of a user's connection to the \systemName network, including signaling for initial authentication and establishing shared security state between the user's device and the iNB to which it connects.
Operators that employ usage-based billing would also deploy their standard Online Charging Service in the \systemName core to support charging.

Conceptually, the SubDB is exactly equivalent to a combined HSS (Home Subscriber Server) and AuC (Authentication Center) in an LTE core network or the UDR (Unified Data Repository), UDM (Unified Data Management) and AUSF (Authentication Service Function) in a 5G core.
Similarly, the SME is analogous to a lightweight MME (Mobility Management Entity) or Access and Mobility Function (AMF) that retains only functionality related to access control and user configuration.
\systemName has no need for a function like the 5G core's Session Management Function (SMF) given our approach to distributing the user plane.

\textbf{Authentication.}
\systemName adopts 3GPP's existing SIM-based authentication scheme, which uses a symmetric secret key provisioned by the network operator into user's SIM cards as a root of trust that is in turn used to derive a series of ephemeral keys while the user is connected to the network.
Although other authentication schemes that align with \systemName's distributed architecture could provide enhanced flexibility~\cite{cellbricks} or privacy~\cite{loca, pgpp}, we reuse the existing 3GPP authentication scheme due to its widespread adoption in existing devices and well-understood security properties.

During initial authentication, the UE performs mutual authentication with the \systemName network in the same manner as done today with 3GPP networks using LTE-AKA.\footnote{We focus on LTE authentication here for simplicity; 5G's authentication protocol is analogous.}
LTE-AKA relies on two key pieces of global state shared between the UE's SIM card and the network (in \systemName, the SubDB): a permanent per-user secret key $K$ and a monotonically increasing sequence number SQN incremented on each authentication attempt to prevent replay attacks.

In addition to these global values used for network authentication, temporary keys used for encryption and integrity protection between the UE and iNB also require global shared context for the duration of a UE's connection to the network.
Every time a UE performs handover to a new RAN element, this $K_{eNB}$ is cycled to provide forward secrecy; the new key is derived from the old key plus a sequence number (NCC) shared between the UE and the SME.

\textbf{Initial attachment.}
\label{s:fm_attach}
To support \systemName compatibility with 4G LTE and 5G UEs, our network attach procedure implements a modified version of the 4G LTE attach protocol.
On attach, \systemName UEs establish radio links with iNBs and send an attach request to the SME.
The SME receives UE attach requests and authenticates the UE as outlined above.
If authentication is successful, derived session keys are distributed to the serving iNB and the SME assigns a unique private IP address to the UE.
As attach completes, the UE sets up a default QoS flow for use for connectivity with the Internet.

\subsection{Edge Control Plane}
\label{s:edge_control_plane}
\systemName introduces an edge control plane for latency-sensitive but ephemeral control signaling that needs to scale with user activity.
The key tasks of this layer are facilitating handover and charging: as users move more frequently or consume more data, the control plane must be able to scale to accommodate this additional load.

\textbf{Handover.}
\label{s:ho_procedure}
Although \systemName assumes mobility-tolerant application and transport layers, coordination of \emph{physical} layer mobility provides a useful performance optimization without violating the end-to-end principle: ideally, the time between when a UE physically disconnects from one iNB and connects to another is minimal.
This requires the target iNB to allocate resources for the incoming UE and to share radio configuration and state information with the incoming UE that will allow it to quickly synchronize to the new iNB's wireless channel.
Thus, \systemName preserves in-network support for coordinated handovers between iNBs, as done today in 3GPP networks.
The critical distinction, however, is that handover \emph{is only a physical layer optimization} and does not involve reconfiguration of the underlying user plane.

Like 3GPP networks, \systemName supports both core-assisted and direct handover, analogous to S1 and X2 handover in LTE.
Because we remove the central user plane, the key distinction between these two handover modes is in handling of security state.
As discussed in Section~\ref{s:central_core}, preserving forward security during handover requires cycling encryption keys used between iNBs and UEs, which in turn requires security state from the central core.
Alternatively, iNBs can \emph{share and reuse} these session keys across iNBs, as done in LTE's X2 handover approach, only cycling keys asynchronously after handover.
In our implementation (\S\ref{s:implementation}), we update these keys on each handover, though this is not required; network operators may choose to trade off control plane load for forward secrecy.

In both cases, iNBs take the same steps to facilitate handovers.
Handover is triggered by the serving iNB, which passes user plane configuration (such as QoS) and security context to the destination iNB.
If using core-assisted handover, this initial message is sent via the SME so it can generate new security keys for the target iNB; otherwise, the source iNB shares its current security key with the target for reuse.
The destination iNB allocates radio state in preparation for receiving the mobile UE, and passes this radio state to the source iNB in the form of a handover command.
The source iNB then forwards the handover command unmodified to the UE, which uses this information to establish a physical link with the destination iNB.
On successful handover completion, target iNB informs source iNB, which releases all UE context.

Unlike 3GPP networks, \systemName does not buffer in-flight traffic towards the UE's old location on the source iNB; we assume that the application or transport layer will recover from any packet loss incurred during handover at a higher layer.
That said, our stateless NAT approach outlined in Section~\ref{s:user_plane} optionally provides a lightweight way to minimize this loss.
After performing handover, the source iNB maintains a "recently moved" table of UE identifiers and target iNB locators.
If any incoming packets arrive with the UE's identifier, the source iNB can simply rewrite the locator portion to match the target iNB's and forwards the packet back into the network.
Entries in such a table need only be held for a short time, as services with which the UE was communicating should learn of its new location within a small number of RTTs.
This forwarding minimizes packet loss for in-flight data during handover without relying on tunnels, per-flow state, or any long-lived state.
We leave implementation and evaluation of this optimization to future work.

\textbf{Handover Proxies (HOPs).} To support scalable and fast handover signaling between iNBs, we introduce Handover Proxies (HOPs).
iNBs use HOPs during handover solely for exchanging RAN-related handover state; in the case of a direct handover, these are the only control plane elements involved in the procedure.
Each iNB connects to one or more HOPs; we envision that network operators will group geographically proximate iNBs into the same HOPs, as handover cannot physically take place between geographically distant iNBs.
Because HOPs are stateless, these can be distributed throughout the network to minimize signaling latency between iNBs and replicated for resilience.

In principle iNBs could communicate directly if all iNBs were directly routeable, making HOPs unnecessary.
However, this may not be possible in practice at scale, as it would require full mesh connectivity between nearby iNBs and more complex configuration and discovery process, as each iNB would need to be configured with the address of any iNB to which it may perform a handover.
HOPs simplify configuration as they do not require direct connectivity between iNBs, allowing operators to change radio topology (adding or removing iNBs) without reconfiguring RAN control plane.

\textbf{Charging Proxies (CPs).}
\label{s:charging_proxies}
Traditional online charging systems allocate a small quota of data to an enforcement function on the data plane.
The enforcement function counts traffic until the quota nears expiration then requests a new quota from the OCS, which can determine whether or not to grant the request.
In principle, iNBs could request quotas from a network operator's central OCS.
However, because \systemName distributes the user plane to the edge, this would incur wide-area network latencies between the OCS and the enforcement function in the iNB.

To alleviate this, iNBs do not directly communicate with the OCS but instead rely upon intermediate Charging Proxies (CPs) that, like HOPs, are deployed close to the network edge and support a number of iNBs.
CPs request quotas from the OCS on behalf of the iNB in batches and maintain these quotas as ephemeral state; in turn, an iNB requests sub-quotas from the CP for each user.
CPs effectively act as caches for the OCS across a range of iNBs, reducing control plane load on the central core while at the same time providing fast responses to iNBs.
\section{Implementation}
\label{s:implementation}

To demonstrate feasibility and to evaluate the performance of our design, we prototype the data and control planes of \systemName by implementing the \systemName SME, HOP, and iNB components.\footnote{We do not implement an OCS or CP in our prototype.}\textsuperscript{,}\footnote{Our implementation of \systemName will be made available.}
Our prototype implements core-assisted handover, which we later compare to the analogous LTE S1 handover in Section~\ref{s:appeval} evaluation of handover impact on QUIC application performance.

\textbf{SME.} We implement \systemName SME attach and handover functionality by removing LTE attach and handover control signaling from the Open5GS \cite{open5gs} MME.

\textbf{HOP.} Our \systemName HOP element is implemented by remove any session UE state in handover control flow from an Open5GS MME.
This allows the HOP to serve as a stateless message-passing proxy during handover.

\textbf{iNodeBs.} We implement iNBs by modifying software srsRAN LTE eNB base stations and modified Open5GS UPFs.
Modified software base stations forward user traffic to a unique modified UPF, which is responsible for passing traffic between the cellular network's private address space and external Internet address space.
eNBs are also modified to pass the appropriate handover control messaging through the HOP and support manually triggered handover.

\textbf{\systemName software testbed.}
Our software testbed is comprised of our SME, a HOP, two iNodeBs, an unmodified srsRAN~\cite{srsran} software LTE UE, and an unmodified HSS used as a SubDB.
Radio links are emulated with srsRAN's ZeroMQ virtual radio and GNU Radio Companion~\cite{gnuradio}.
We run the UE and the radio interface half of the iNodeBs on a single x86 laptop (i7-11800H @ 2.30GHz CPU and 16GB RAM).
The user plane portion of each iNodeB are deployed on their own x86 laptops.
The HOP is deployed on a less powerful x86 PC equipped with an Intel J3060 CPU (1.60GHz) and 8GB of RAM and connect it to the iNodeBs.
Our central core network is composed of one SME and one SubscriberDB instance, both of which are deployed on an x86 VM.

\textbf{LTE testbed.} To benchmark \systemName against a comparable LTE network, we construct an comparable LTE testbed from standard srsRAN LTE eNodeBs connected to an Open5GS EPC instance running on an x86 VM.
We modify the eNodeB implementation only to allow triggering handover programatically; the RAN and core are otherwise unmodified open source implementations of a standard 3GPP LTE network.

\begin{figure}
  \centering
  \includegraphics[width=0.7\columnwidth]{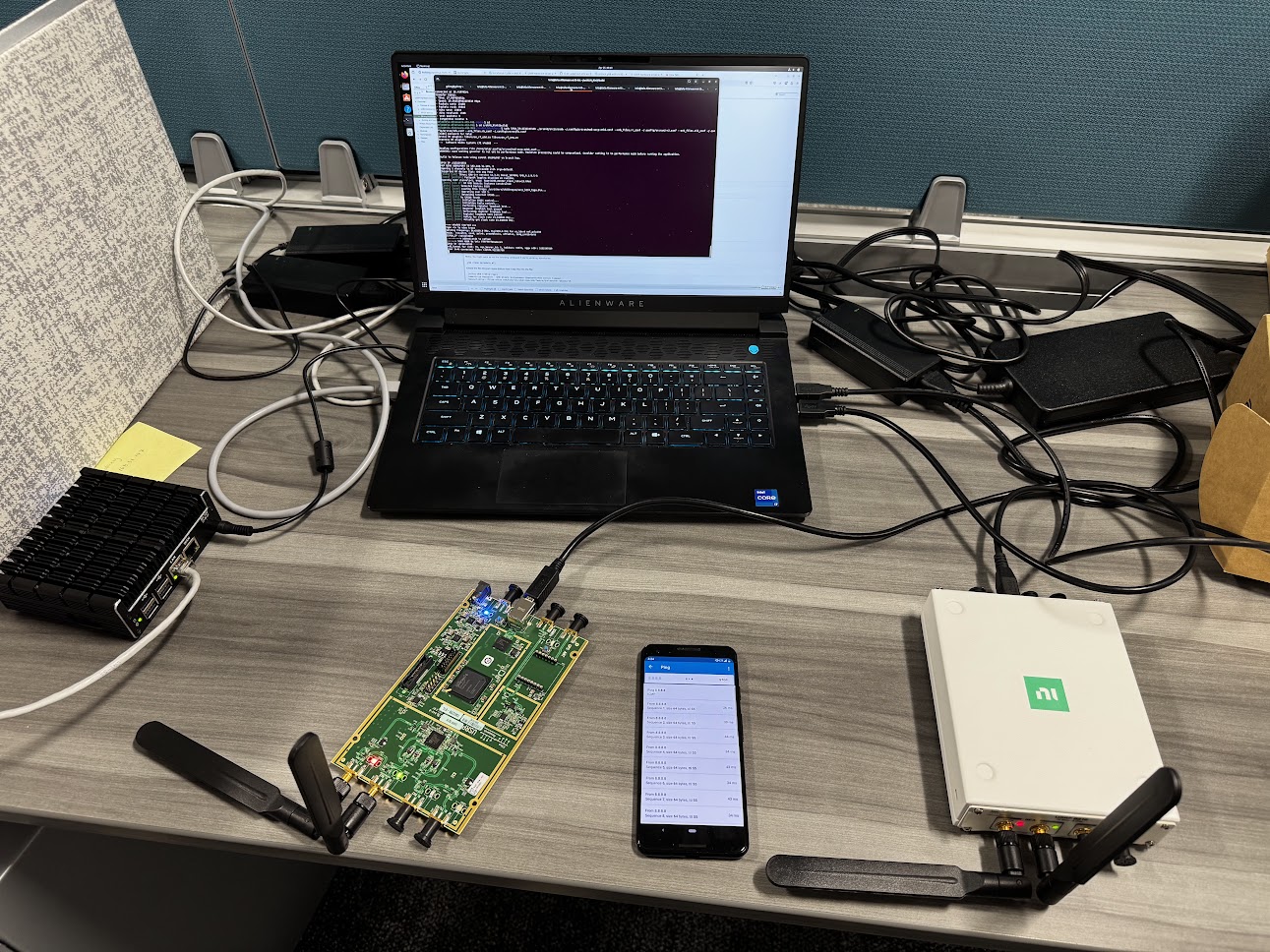}
  \caption{\systemName over-the-air testbed.}
  \label{fig:deployment}
\end{figure}

\textbf{Over-the-air testbed.}
To evaluate unmodified commercial applications on \systemName and LTE, we replace the virtual radio interface with B210 software radios and replace the software UE with an unmodified Google Pixel 3 (Figure~\ref{fig:deployment}).

\section{Evaluation}
\label{s:evaluation}

We first validate our assumptions about end-to-end mobility support on today's Internet (\ref{s:appeval}).
We then quantify \systemName reduction of latency and control plane overhead (\ref{s:eval:latency}, \ref{s:eval:scaling_cp}, \ref{s:eval:packet_loss}).

\subsection{\systemName Handover Performance Impact}
\label{s:appeval}
To be viable, \systemName must support applications without modification. 
We evaluate this using both custom QUIC-based applications and unmodified commercial applications.
Without access to commercial infrastructure and traffic we are unable to evaluate performance impacts of \systemName mobility handling under realistic user traffic loads and mixes.
We focus our evaluation on benchmarking performance impacts to individual applications to provide a basis for reasoning about mobility impacts on commercial traffic.

\begin{table*}[htbp] 
\centering
\caption{QUIC toy applications and metrics used to characterize handover performance in LTE and \systemName.}
\label{t:app_metrics_comparison}

\begin{tabularx}{\textwidth}{@{\extracolsep{\fill}} l l l c c c c} 
\toprule
% Headers spanning multiple columns
\textbf{Application} & \textbf{Performance Metric} & \textbf{Unit} & \multicolumn{2}{c}{\textbf{LTE}} & \multicolumn{2}{c}{\textbf{\systemName}} \\
% Mid-rules to group the columns under LTE and \systemName
\cmidrule(lr){4-5} \cmidrule(lr){6-7} 
% Sub-headers for Handover (H) and No Handover (NH)
& & & \textbf{w/ HO} & \textbf{w/o HO} & \textbf{w/ HO} & \textbf{w/o HO} \\ 
\midrule
% Data rows - using \multirow for Application name spanning multiple metric rows
\multirow{1}{*}{{Bulk File Transfer}} & Average throughput & Mbps & 26.1 $\pm$ 3.0 & 26.9 $\pm$ 1.0 & 26.7 $\pm$ 3.2 & 30.5 $\pm$ 0.9 \\
\addlinespace 
\hline

\multirow{1}{*}{{Live Video Streaming}} & Playback Framerate & fps & 24.2 $\pm$ 0.3 & 24.2 $\pm$ 0.5 & 24.2 $\pm$ 0.3* & 24.2 $\pm$ 0.1 \\ 
\addlinespace
\hline

\multirow{3}{*}{{Buffered Video Stream}} & Buffer stall time & s & 0 & 0 & 0 & 0 \\
& Average playback quality & (1-5) & 5 & 5 & 5 & 5 \\
& Average buffer length & s & 39.2 $\pm$ 0.5 & 39.2 $\pm$ 0.3 & 38.7 $\pm$ 0.5 & 38.5 $\pm$ 0.3 \\
\bottomrule
\end{tabularx}
\end{table*}

\textbf{QUIC applications.}
We first build and instrument three toy applications to isolate and quantify the effects of \systemName handover on the performance of applications that use QUIC transport.
We select applications domains (bulk file transfer, buffered video streaming, and live video streaming) that are sensitive to mobility; results are shown in Table~\ref{t:app_metrics_comparison}.

\textit{Bulk file transfer.}
We implement a simple file transfer application using quinn~\cite{quinn}, an open source QUIC implementation.
Our application downloads a single 100MB file from an external web server and records average throughput.
We evaluate in both non-mobile and mobile scenarios, triggering one handover event per download, and repeating 50 times.
Although average throughput is lower in mobility scenarios for both LTE and \systemName, performance is comparable, with \systemName slightly outperforming LTE in the stationary case.

\textit{Buffered video streaming.}
Our buffered video streaming application requests chunks of video (also using quinn~\cite{quinn}), implementing a simple buffer-based adaptive bitrate algorithm~\cite{abr_buffer} that requests video chunk quality based on instantaneous buffer-length.
We find buffered video streaming performance, measured by stall rate and average buffer length, is identical in all cases.

\textit{Live video streaming.}
We measure the impact of handover on live video streaming by instrumenting an implementation of a Media Over QUIC (MoQ) live video client.%toy application
MoQ is a QUIC live media application protocol currently undergoing standardization by the IETF~\cite{ietf-moq-transport-10}.
We find that \systemName handovers cause application failures in our MoQ implementation.

Closer inspection revealed that QUIC's passive approach to IP mobility handling causes application failure in one mobility edge case.
QUIC's connection migration feature employs a passive mechanism for handling IP address changes.
QUIC permits client endpoints to change IP addresses freely, with server endpoints only updating a client's location upon receiving a packet from the new IP address with a recognized Connection ID.
With this passive approach, connections may fail to complete (consequently causing application failure) if the mobile client has no data to send, leaving the server unaware of the client's new IP address.

We encounter this issue with QUIC in our MoQ application due to the nature of the subscriber traffic pattern.
If the QUIC client changes IP address between the delivery of live video frames (which is the majority of the connection's lifetime), the QUIC client will never send a packet alerting the server of its new address. 
To address this, we propose a simple app-level fix: given that the subscribed MoQ client expects a constant flow of live video frames from the MoQ server, the client should infer that mobility has occurred when it does not receive a live video frame at the expected time and should send a ping to the MoQ server to alert it of its new IP.
We implement this logic in the MoQ app by modifying Chromium's QUIC library and build it into the browser.
After modifying Chromium we observe that our Media-over-QUIC application no longer fails during \systemName handovers -- indeed, application performance is unaffected by the handover event compared to the LTE and \systemName no-handover baselines (Table~\ref{t:app_metrics_comparison}).

\noindent
\textbf{Unmodified applications.}
We next evaluate unmodified commercial applications to explore how end-to-end mobility tolerance is currently exercised in widely-adopted use cases.
We evaluate \systemName handover on nine popular commercial applications: Zoom, Google Meet, YouTube, Spotify, Facebook, Instagram, Snapchat, Google Chrome, and NewPipe.
We run each application on an unmodified Google Pixel 3 connected to our over-the-air testbed.

\textit{Closed-source applications.}
While we cannot internally instrument proprietary closed-source commercial applications, we observed consistent externally visible behaviors across \systemName handover that indicate these third-party applications tolerate address changes without requiring network-layer anchoring.
Even applications that used TCP functioned as expected across \systemName handovers, indicating broad implementation of application-layer mobility support.

For video calling and streaming media applications, we capture packets from each application and rely on previous work that identifies network usage of these various applications~\cite{webrtc, meet_rtp, zoom_traffic, youtube_dash} to identify packet flows transporting application data (rather than ads or performance monitoring).
We supplement this data with manual observation of user-level application performance.
We find that \systemName and LTE handover had similar performance impacts on video call applications, introducing average jitter of $\approx$14ms/9ms(\systemName) and $\approx$17.5ms/8ms (LTE) for DL/UL traffic on Google Meet and $\approx$11.5/15.5 ms (\systemName) and $\approx$11ms/ 15ms (LTE) on Zoom.
We also see that audio and video playback of streaming media on Spotify and Youtube is uninterrupted by both types of handovers, requiring no re-buffering.
Finally, by correlating Youtube playback bitrate to application data throughput, we see handover type has no effect on adaptive bitrate-determined playback quality.
Table \ref{t:real_app_performance} summarizes our results.

For social media and web browsing applications, similar fine-grained analysis of application performance is complicated due to the small sizes of application data and short (<100ms) resolution time of network requests.
This made (1) distinguishing encrypted application data packet flows from encrypted ad or application monitoring packet flows and (2) consistently triggering handover during ephemeral network requests difficult.
To evaluate these application types, we define construct well-defined, repeatable test cases that verify whether sequences of application interactions are robust to \systemName handovers: for each application interaction, we verify whether a series of interactions, interrupted with handover upon interaction completion, can be performed without additional user input.
Appendix \ref{s:appendix:app} further outlines these test cases.
We find that \systemName and LTE handover between interactions has no impact on subsequent functionality.

\textit{Open-source application.} The sole open-source commercial application evaluated was NewPipe, a video streaming app that requests content directly from YouTube's commercial servers (though due to limited API permissions, NewPipe only supports non-adaptive single bitrate video streaming.) 
We find that NewPipe's internal buffer sizes and playback stall rates are unaffected by handover type -- indicating that current commercial YouTube video server deployments widely support client device IP mobility and suggesting that \systemName has minimal impact on buffer-based adaptive bitrate algorithms.

\begin{table*}[htbp]
\centering
\caption{Performance evaluation of popular media streaming applications}
\label{t:real_app_performance}
\begin{tabularx}{\textwidth}{@{\extracolsep{\fill}} l l c c l l}
\toprule
\textbf{\multirow{2}{*}{Application}} & \textbf{\multirow{2}{*}{Interaction}} & \textbf{\multirow{2}{*}{QUIC Protocol}} & \multicolumn{3}{c}{\textbf{\makecell{Performance Impact}}} \\
\cmidrule(lr){4-6}
\textbf{} & \textbf{} & \textbf{} & \textbf{Metric} & \textbf{LTE} & \textbf{\systemName} \\
\midrule
\multirow{2}{*}{Google Meet} & \multirow{2}{*}{Video Call} & \multirow{2}{*}{N}  & Jitter DL (ms) & 17.5 $\pm$ 1.8 & 14.0 $\pm$ 0.5 \\
& & & Jitter UL (ms) & 8.0 $\pm$ 2.0 & 8.7 $\pm$ 1.4 \\
\addlinespace
\hline
\addlinespace
\multirow{2}{*}{Zoom} & \multirow{2}{*}{Video Call} & \multirow{2}{*}{N} & Jitter DL (ms) & 10.8 $\pm$ 0.1 & 11.6 $\pm$ 0.4 \\
& & & Jitter UL (ms) & 14.9 $\pm$ 1.8 & 15.5 $\pm$ 2.5 \\
\addlinespace
\hline
\addlinespace
\multirow{4}{*}{Youtube} & \multirow{2}{*}{Buffered Video Playback} & \multirow{2}{*}{Y}  & Stall Rate & 0 & 0\\
& & & Video Quality & 720p (max) & 720p\\
\addlinespace
& \multirow{2}{*}{Live Video Playback} & \multirow{2}{*}{Y} & Stall Rate & 0 & 0\\
& & & Video Quality & 720p (max) & 720p\\
\addlinespace
\hline
\addlinespace
Spotify & Music Playback & N & Stall Rate & 0 & 0\\
\addlinespace
\hline
\addlinespace
\multirow{2}{*}{NewPipe*} & \multirow{2}{*}{Buffered Video Playback} & \multirow{2}{*}{N}  & Avg. Buffer Size (s) & 52.6 $\pm$ 0.16 & 52.5 $\pm$ 0.20 \\
& & & Stall rate & 0 & 0\\

\addlinespace
\bottomrule
\end{tabularx}
\end{table*}

\noindent\fbox{
    \parbox{0.95\columnwidth}{
        {\textbf{Takeaway:} Today's transports and applications support end-to-end mobility. QUIC-based applications are minimally affected by \systemName handover.
        In one edge case, \systemName handover can trigger deadlock, but this can be resolved with a single-packet ping.
        Real applications are minimally affected by \systemName handover regardless of QUIC use.}
    }
}

\subsection{Latency Gains from Edge Routing}
\label{s:eval:latency}

We next turn to quantifying the end-to-end latency gains that \systemName achieves by avoiding detouring traffic through IP anchor points.
We rely on simulating gains built from real infrastructure deployment data in this evaluation, as evaluating gains on a commercial network deployment is beyond the resources of an academic institution.
Even evaluating these gains in partnership with a commercial mobile network operator would require high capital investments akin to deploying a NextG network.
Consider a scenario in which a network operator seeks to maximize the number of their users that are able to reach a set of servers outside their network (e.g., a CDN) within a given latency budget.
In 3GPP networks today, an operator's ability to maximize this is constrained by the placement of their IP anchors (P-GWs), which drives the degree of latency inflation users experience due to detouring.
The only way to reduce this latency inflation is by adding more IP anchors to their network.
At the same time, latency is also bounded by the number and placement of peering PoPs of the operator's IP network: if all IP traffic must traverse a single router before reaching the CDN, then increasing the number of IP anchor points in the cellular network will not make a difference.
Intuitively, because \systemName avoids IP anchors we expect that it provides lower latency service than a 3GPP network given any particular distribution of PoPs.

\begin{figure*}[t!]
  \centering
  \begin{subfigure}[t]{0.45\textwidth}
    \centering
    \includegraphics[width=0.9\columnwidth]{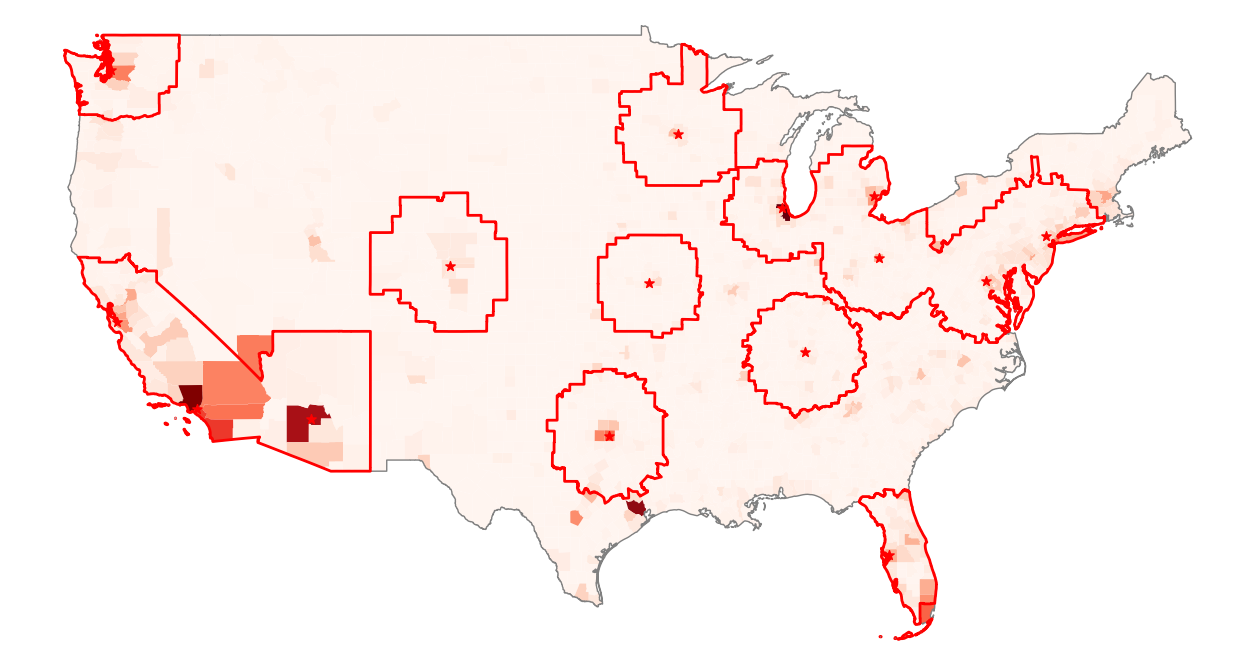}
    \caption{15 core deployment with latency budget of 300km.} %300km latency cutoff
  \label{fig:latency_map}
  \end{subfigure}
  ~
  \begin{subfigure}[t]{0.45\textwidth}
  \centering
  \includegraphics[width=0.7\columnwidth]{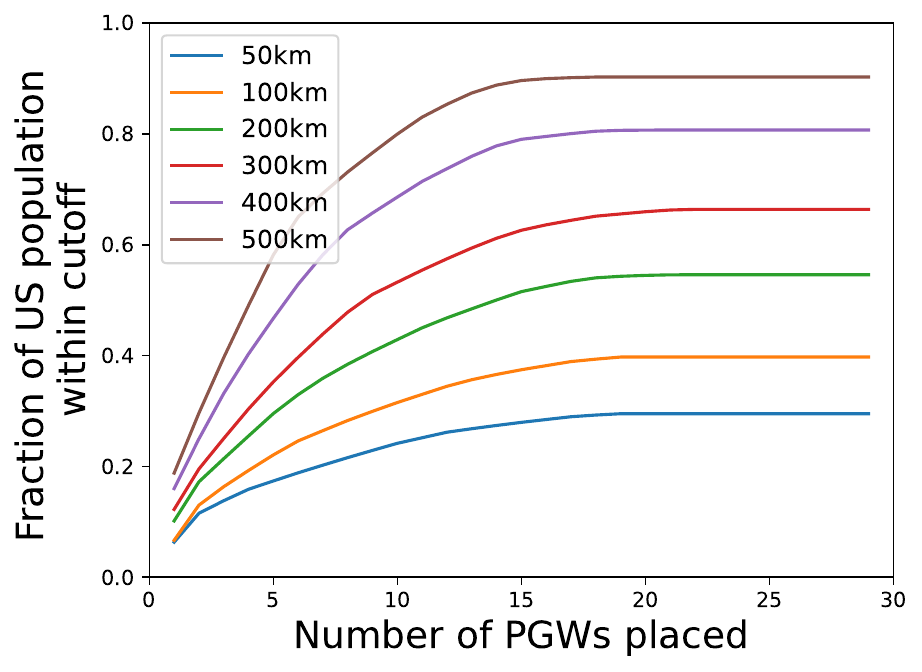}
  \caption{US population covered by varying core/latency budgets}
  \label{pop_coverage}
  \end{subfigure}%
  \caption{
  Percentage of the US population that an MNO with access to AT\&T's PoP footprint could serve within a given latency budget and varying 3GPP core network location budget.
  We find 95\% of latency gains can be achieved by deploying approximately 15 cores, though this leaves 47\% of the population unable to meet a latency target of 300km. The map depicts an example layout given a latency threshold of 300 km and 15 cores.}
\end{figure*}

To explore this tradeoff, we adopt an analytical approach based on publicly-available data on peering locations, CDN footprints, and US county population to quantify the latency improvement of \systemName compared to current cellular network deployments.
For each county, we compute the shortest possible distance connecting the centroid of the county, a cellular core location, a network operator's peering point, and a CDN location.
We then set a fixed budget of core networks it can deploy and determine what percentage of the US population could be served under a given latency budget.
For mobile core placement, we adopt a greedy deployment strategy in which network operators deploy mobile cores to maximize the total US population they can serve under a fixed latency/distance cutoff.
Note that this establishes a generous baseline for comparison as such optimal placement may not be possible in practice.
Figure \ref{fig:latency_map} illustrates this deployment strategy for a core budget of five and a "latency" budget of 300km.
We then make the simplifying assumptions that (1) mobile network operators deploy network cores at peering points in their network (another generous assumption, as doing otherwise introduces unnecessary additional latency to user data path) and (2) latency is directly proportional to the straight-line distance between two points.
To ground our analysis in a realistic deployment, we use publicly-available datasets of US county populations~\cite{us_census}, AT\&T peering points from PeeringDB~\cite{peering_db},  and Cloudfront CDN PoP locations~\cite{cflare_pop}.

Figure~\ref{pop_coverage} illustrates that increasing a network operator's network core "budget" -- i.e., the number of P-GWs deployed -- increases the proportion of the total US population within a fixed distance cutoff but faces diminishing returns.
The best possible coverage a mobile network operator can achieve under the 3GPP model occurs when the operator deploys a mobile core at each peering point, as every user's traffic will be routed to the Internet through the closest peering point.
\textit{Note that with \systemName, network operators achieve the optimum "for free" without deploying any additional infrastructure on top of their existing IP network.}

To estimate financial costs associated with \systemName's more attractive scaling economics, we use data from the FCC's Catalog of Eligible Expenses~\cite{FCCCostCatalog2021}, which captures the installation costs of critical network infrastructure such as cellular core sites and border routers.
For the nation-scale networks we consider, adding a new core site costs on average \$2.75M.
In comparison, deploying redundant 100G border routers to an existing PoP costs on the order of \$200,000, including hardware and deployment costs.
This means that to achieve only 80\% of the benefits of \systemName, a network operator with AT\&T's PoP footprint would need to invest over \$27M in ten core network sites compared to just \$6.6M to deploy high-end border routers at all existing PoP locations.
Put another way, a ten PoP \systemName deployment matches 3GPP networks at a cost of only \$2M, representing over 90\% lower costs.

Beyond financial costs, provisioning more network cores to realize these latency gains comes with a significant inflation in control plane overhead. 
In a multiple core deployment, the network control plane must handle additional control messaging if a device's "serving core" changes during handover~\cite{3gpp23502ssc_handover, 3gpp23502session_management}, as outlined in the description of 5G MEC in \S~\ref{s:related}.

\begin{figure}
  \centering
  \includegraphics[width=0.3\textwidth]{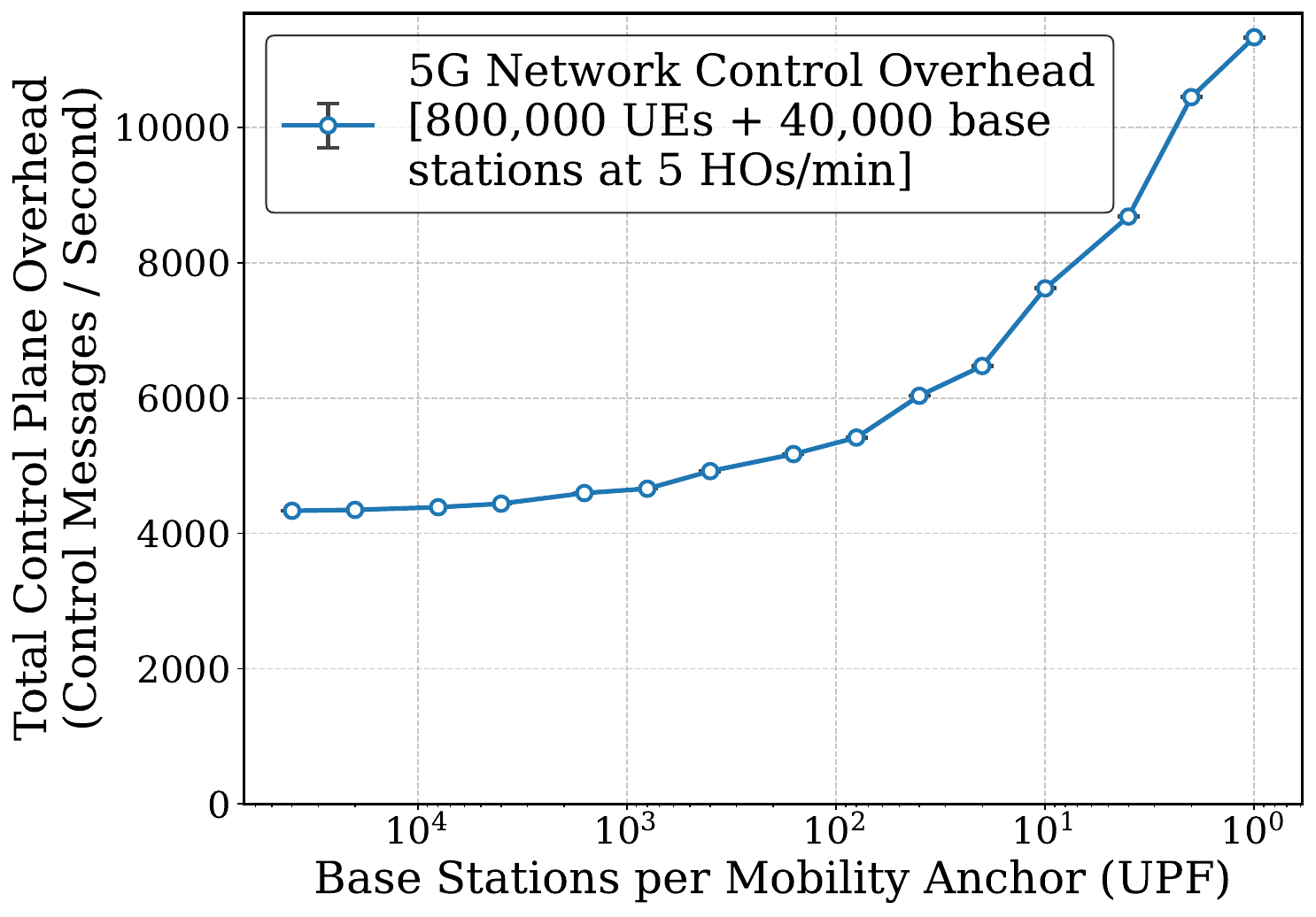}
  \caption{Control plane scaling in a 5G MEC simulation.}
  \label{fig:mec_scale}
\end{figure}

Figure \ref{fig:mec_scale} illustrates how such control message overhead in increasingly dense 5G MEC deployments scales.
We simulate mobility in a 3GPP network by modeling the network as a grid in which 800,000 mobile devices are distributed across 40,000 base stations and handover between base stations probabilistically at an average rate of 5 handovers per minute.
We then vary core/mobility anchor deployment density in our simulation from a minimal deployment in which a single anchor handles all user traffic in the network to a hyper-dense deployment in which each base station is served by its own anchor (the configuration required to match \systemName performance).
We find that matching the latency gains of an \systemName deployment incurs up to a 3.3x increase in control messaging in equivalent dense anchor deployments.

\noindent\fbox{
    \parbox{0.95\columnwidth}{
        \textbf{Takeaway:} By removing in-network mobility, \systemName can reduce the financial cost of providing low latency service to users by more than 90\% while reducing additional control message overhead by 70\% compared to 3GPP networks.
        
    }
}

\subsection{Scaling Gains of Simpler Control Plane}
\label{s:eval:scaling_cp}
In addition to lowering end-to-end latency, removing the user plane from the network core reduces coordination of network state across the edge and core networks during handover.
This translates to reduced control plane load during \systemName handover during normal operation.

We demonstrate the implications of this control message reduction on overall system performance by observing the scaling properties of our system.
We scale up our software testbed to connect 32 UEs and 64 base stations deployed across two 62GB, 55-core x86 servers and restrict CPU cycles available to the machine running the \systemName core to push the hardware close to operational capacity.
We take this approach to evaluation due to lack of access to commercial network infrastructure and commercial simulation equipment.
Although commercial simulation equipment exists that would allow simulations of higher handover rates, we do not have access to this equipment due to cost.
Our testbed is limited to 32 UEs across 64 base stations due to the constraints of hardware and open source implementations of the LTE/5G core and radio interface emulator.

\begin{figure}
  \centering
  \includegraphics[width=0.3\textwidth]{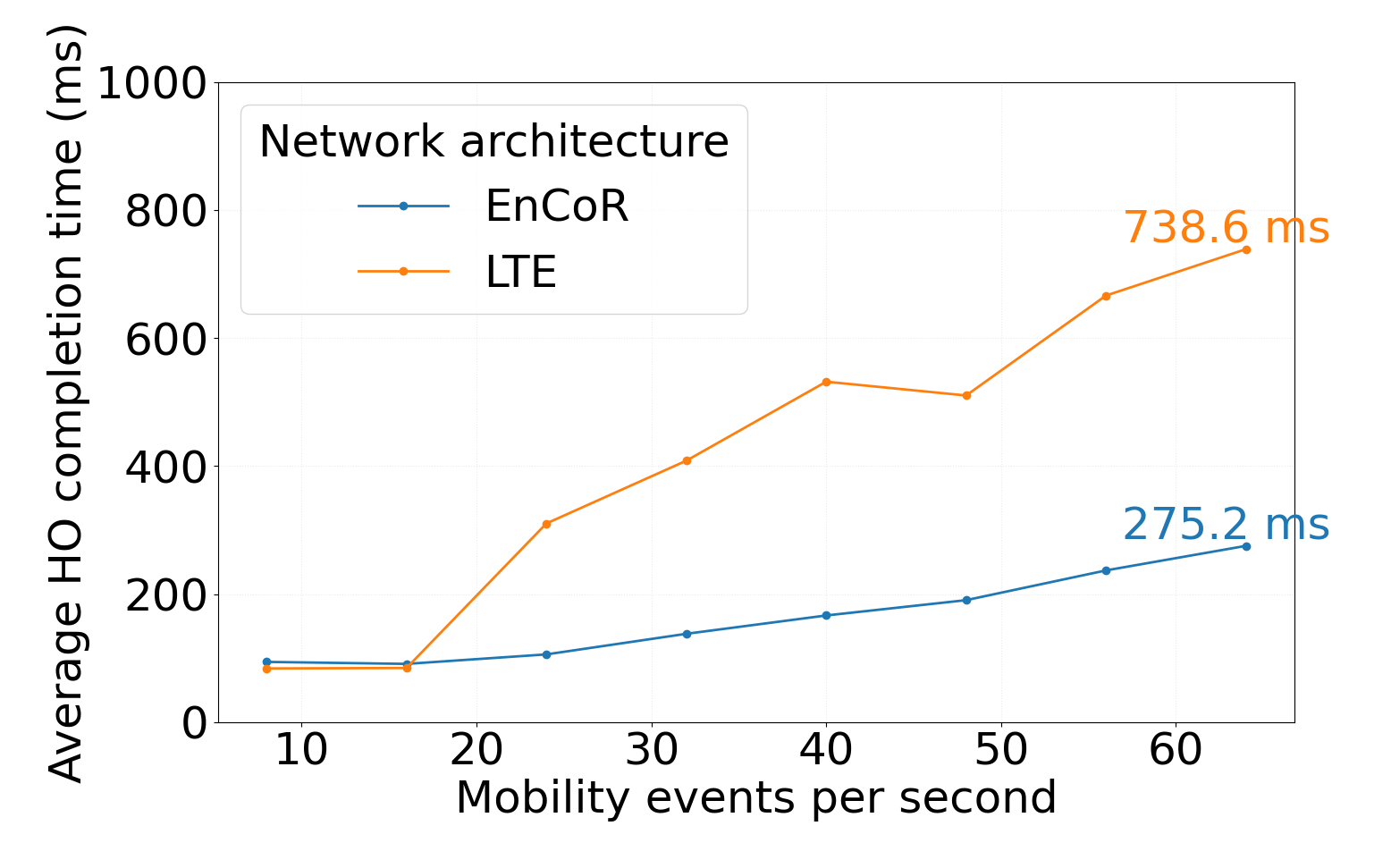}
  \caption{Handover completion times under load. \systemName achieves improvements of up to 2.6x faster completion in a resource-constrained deployment.}
  \label{fig:scale}
\end{figure}

Figure \ref{fig:scale} shows the impact of increasing the handover load offered by these devices on handover completion time.
As the number of handovers per second in both networks increase, handovers in both LTE and \systemName take longer to complete.
However, because the \systemName control plane load is lighter during handover under equivalent conditions, \systemName handover latency is less sensitive to increases in mobility load.
At a peak of 64 mobility events per second, \systemName handovers complete in an average of 275.2 ms, while LTE handovers complete in an average of 738.6 ms, a 2.6x reduction.

Further, \systemName achieves these gains without simply shifting the burden of mobility control messaging to content providers.
QUIC's current approach to supporting IP mobility introduces at most 2 additional control signaling packets, which drops to 0 when the control frames are included in already scheduled packets~\cite{ietf_quic}.
\S \ref{s:appeval} revealed that QUIC's server-side implicit detection of client mobility can cause connection failures in certain edge cases, which we propose resolving by requiring applications to send a "ping" packet upon \systemName mobility -- requiring at most 1 additional QUIC control packet for complete end-to-end mobility support.
Table \ref{t:ctrl_msg_eval} summarizes these findings.

\noindent\fbox{
    \parbox{0.95\columnwidth}{
        \textbf{Takeaway:} \systemName reduces mobility control overhead, allowing it to handle more mobility with less performance degradation than LTE.
        We demonstrate performance gains of up to 2.6x in handover completion time, achieved without pushing mobility signaling overhead to other layers.
        }
}

\begin{table}
\begin{tabular}{|| m{3.1cm} | m{0.8cm} | m{1.5cm} | m{0.8cm} || } 
 \hline
 {} & {QUIC} & {Network (via core)} & Total \\
 \hline\hline
 LTE & 0  & 15 (15) & 15\\
 \hline
 \systemName & < 2 & 7 (2) & 7 - 9 \\
 \hline
 \systemName + mod. QUIC & 1 - 3 & 7 (2) & 8 - 10 \\
 \hline
\end{tabular}
\caption{Control messages required for handover}
\label{t:ctrl_msg_eval}
\end{table}

\subsection{Packet Loss From Implicit Mobility}
\label{s:eval:packet_loss}
While QUIC's implicit approach to mobility signaling adds minimal additional control signaling to existing QUIC connections, it creates a window of opportunity in which packets may be mis-addressed and sent to an outdated client location.
During the period of time between the mobile client's movement to a new IP address and the static server's reception of a packet indicating the mobile client's new IP, there is a possibility that an application server continues to send packets through the network to the client's stale address.

The number of these "doomed" packets varies depending on application type and state: in the best case, the QUIC client endpoint has received all in-flight packets from the server when mobility occurs, and no packets are lost.
In the worst case, the server is in the middle of sending a large chunk of data before receiving a location update, and a large chunk of its in-flight packets are delivered to the wrong address.

We briefly explore the strictly worse-case dynamics of lost packets during handover on our \systemName prototype testbed by measuring the downlink packet retransmission rate of a bulk file transfer interrupted by a handover event. 
We find that Bulk File Transfer over \systemName sees a 130\% increase in retransmission rate compared to the LTE baseline.
When we re-run this experiment with the buffered video application, which requests video chunks periodically rather than continuously, we see only a 15\% increase in retransmission rate compared to the LTE baseline.
At the same time, despite this increase in retransmission application level metrics such as throughput remain comparable between the two.

We note that even the worst case scenario is easily addressed with a straight-forward performance optimization: source iNBs forward mis-addressed UE traffic to the UE's new address at the target iNB, which can be elegantly achieved in \systemName due to the "stateless NAT"-based approach to UE IP addressing as discussed in Section~\ref{s:edge_control_plane}.

\noindent\fbox{
    \parbox{0.95\columnwidth}{
        \textbf{Takeaway:} Although QUIC's implicit approach to mobility incurs modest increased packet loss in \systemName, application performance appears unaffected. \systemName can mitigate this loss with simple, tunnel-free forwarding enabled by its transparent IP assignment approach. We leave implementing this optimization to future work.
    }
}
\section{Discussion}    
\label{s:discussion}
\noindent
\systemName's adoption is enabled by widespread QUIC usage and application-layer mobility support.
Our evaluation of real apps builds upon previous work~\cite{meta_quic, akamaiquic, amazonquic, cloudflarequic, facebook_quic_adoption_2020, 10.1145/3727063.3727066} that suggests these mechanisms are widely present today's Internet.

However, instag of traffic on today's Internet is not mobility tolerant.
Supporting legacy traffic is straightforward in \systemName.
We envision deployment in the same manner as other generational upgrades to cellular networks in the past: initially in parallel to existing infrastructure until legacy systems can be decommissioned.
Thus, iNodeBs can direct legacy traffic to existing legacy core network infrastructure that provides in-network mobility support.
For example, any TCP traffic could be re-routed to the legacy core by an iNB.
As more traffic migrates to QUIC or other mobility tolerant transport protocols, legacy cores can be scaled down or removed entirely.

Quantifying the overhead of this integration process in engineering hours or monetary expense is difficult even for operators deploying commercial systems today.
We instead characterize integration overhead in terms of engineering scope and architectural disruption.
EnCoR requires simple changes to the base station's user plane, removes much of the core's control plane, introduces 2 new stateless components that can be deployed at scale on commodity hardware, and leaves the cellular air interface and user devices untouched.
Our prototype required under 1650 additional lines of new user-space code and completely removed 1 control plane interface and associated state machine, in addition to removing handling of 13 control message types from another control plane interface, at the MME, on top of completely removing LTE S-GW and P-GW components.
This indicates that adopting \systemName requires substantially fewer modifications than previous generational transitions, which required coordinated changes across user devices, radio access network, and core.
\section{Conclusion}
\label{s:conclusion}
From the first commercial cellular networks, in-network support for mobility has been an unstated requirement.
The emergence of mobility-tolerant applications and transport protocols invites the community to revisit this assumption.
\systemName demonstrates an end-to-end approach to mobile networks is feasible and has significant benefits in performance, cost, and simplicity compared to traditional 3GPP approaches. 
Despite this radical change to the service model, we demonstrate that \systemName is deployable today alongside existing networks and with unmodified user devices.
In doing so, we show that revisiting core mobility assumptions offers a practical path towards simpler, lower-latency, and more scalable cellular networks.

\bibliographystyle{plain}
\bibliography{reference}

\appendix
\section{Application Performance}
\label{s:appendix:app}

We evaluated seven unmodified commercial applications to understand how \systemName impacted performance.
Some of these applications used the QUIC protocol, while others did not.

We find that \textbf{all real applications we tested supported \systemName handover} -- even those that did not use QUIC for application traffic, indicating that these applications implement some form of \textbf{application-layer mobility support} to restore connections broken by the change of IP address incurred by \systemName mobility.
It seems likely this is because applications must support similar IP address change events associated with switching between Wi-Fi networks or between Wi-Fi and cellular networks.

We then examine the performance impact of \systemName handover for various interactions with these applications.
We note that some app interactions only request small amounts of data from the network at infrequent intervals while other interactions request large amounts of data from the network.
For each of the former interaction types (\textit{Facebook/Instagram profile search}; \textit{Snapchat sending/receiving Snaps}; \textit{Google Chrome web browsing}; \textit{ChatGPT querying}), we simply indicate whether a series of repeat interactions is affected by an \systemName handover in between individual interactions.
Across all such interactions we find that \systemName handover does not affect functionality of subsequent interactions.

For interactions that request large amounts of data from the network (live video calling and video streaming), we calculate relevant statistics from packet traces or record application metadata that allow us to quantify the performance impact of \systemName handover. We find:

\textit{Google Meet and Zoom calling.} \textbf{Video and voice calling interrupted by \systemName and LTE handover incurs comparable performance impacts}.
When we interrupt a Google Meet call with a \systemName handover, we see network jitter of approximately 14ms and 9ms for downlink (DL) and uplink (UL) traffic over a observation period of one minute.
We see that LTE handover has a similar performance impact, introducing average jitter of $\approx$17.5ms/8ms for DL/UL traffic for the same setup. 
For Zoom, we see network jitter of $\approx$11.5/15.5 ms for DL/UL traffic during \systemName handover, and $\approx$11ms/15ms during LTE handover.

\textit{Youtube video and Spotify audio streaming. } We first observe the impact of handover on video and audio playback stall rate, looking at both live and non-live video playback.
We see that \textbf{both \systemName and LTE handover have no impact on stall rate} due to the robust buffers built up by video and audio playback clients, even when streaming at the maximum quality. 
We then observe the impact of handover on video playback quality selected by Youtube's automatic bitrate selection algorithm.
We find that \textbf{video playback quality is unaffected by both LTE and \systemName handover}.  

\begin{table*}[htbp]
\centering
\caption{Performance evaluation of real applications in \systemName.}
\label{t:real_app_performance}
\begin{tabularx}{\textwidth}{@{\extracolsep{\fill}} l l c c l l l}
\toprule
\textbf{Application} & \textbf{Interaction} & \textbf{\makecell{QUIC \\ Protocol}} & \textbf{\makecell{EnCoR \\ mobility \\ Support}} & \multicolumn{3}{c}{\textbf{\makecell{Performance \\ Impact}}} \\
\cmidrule(lr){5-7}
\textbf{} & \textbf{} & \textbf{} & \textbf{} & \textbf{Metric} & \textbf{LTE} & \textbf{\systemName} \\
\midrule
\multirow{2}{*}{Google Meet} & \multirow{2}{*}{Video Call} & \multirow{2}{*}{N} & \multirow{2}{*}{Y} & Jitter DL (ms) & 17.49 & 14.04 \\
& & & & Jitter UL (ms) & 8.05 & 8.73 \\
\addlinespace
\hline
\addlinespace
\multirow{2}{*}{Zoom} & \multirow{2}{*}{Video Call} & \multirow{2}{*}{N} & \multirow{2}{*}{Y} & Jitter DL (ms) & 10.80 & 11.58 \\
& & & & Jitter UL (ms) & 14.91 & 15.47 \\
\addlinespace
\hline
\addlinespace
\multirow{4}{*}{Youtube} & \multirow{2}{*}{Buffered Video Playback} & \multirow{2}{*}{Y} & \multirow{2}{*}{Y} & Stall rate & 0 & 0 \\
& & & & Video Quality & 720p (max) & 720p \\
\addlinespace
& \multirow{2}{*}{Live Video Playback} & \multirow{2}{*}{Y} & \multirow{2}{*}{Y} & Stall rate & 0 & 0 \\
& & & & Video Quality & 720p & 720p \\
\addlinespace
\hline
\addlinespace
Spotify & Music Playback & N & Y & Stall Rate & 0 & 0 \\
\addlinespace
\hline
\addlinespace
Facebook & User/Page Search & Y & Y & \multicolumn{3}{c}{\textit{Functionality Works}} \\
\addlinespace
\hline
\addlinespace
Instagram & User/Page Search & Y & Y & \multicolumn{3}{c}{\textit{Functionality Works}} \\
\addlinespace
\hline
\addlinespace
\multirow{2}{*}{Snapchat} & Sending Snap & Y & Y & \multicolumn{3}{c}{\textit{Functionality Works}} \\
& Receiving Snap & Y & Y & \multicolumn{3}{c}{\textit{Functionality Works}} \\
\addlinespace
\hline
\addlinespace
\multirow{2}{*}{Google Chrome Browser} & Web browsing (Google Search) & Y & Y & \multicolumn{3}{c}{\textit{Functionality Works}} \\
& Web browsing (TCP) & N & Y & \multicolumn{3}{c}{\textit{Functionality Works}} \\
\addlinespace
\hline
\addlinespace
ChatGPT & Asking Questions & N & Y & \multicolumn{3}{c}{\textit{Functionality Works}}\\
\bottomrule
\end{tabularx}
\end{table*}

\end{document}